\documentclass[aps,preprint,nofootinbib,floatfix,showpacs,superscriptaddress]{revtex4}
\usepackage{epsfig}
\usepackage{graphicx}
\usepackage{multirow}
\usepackage{latexsym,amsbsy,amsmath}
\usepackage{stackrel}
\begin{document}
\title{Doubly charmed multibaryon systems}
\author{H.~Garcilazo} 
\email{humberto@esfm.ipn.mx} 
\affiliation{Escuela Superior de F\' \i sica y Matem\'aticas, \\ 
Instituto Polit\'ecnico Nacional, Edificio 9, 
07738 Mexico D.F., Mexico} 
\author{A.~Valcarce} 
\email{valcarce@usal.es} 
\affiliation{Departamento de F\'\i sica Fundamental,\\ 
Universidad de Salamanca, E-37008 Salamanca, Spain}
\date{\today} 

\begin{abstract}
We study two- and three-baryon systems with two units of charm
looking for possible bound states or resonances.
All two-baryon interactions are consistently derived from a 
constituent quark model tuned in 
the light-flavor hadron phenomenology: spectra and interactions. 
The presence of the heavy quarks makes the two-body interactions simpler 
than in the light-flavor sector. Our results show a narrow 
two-body resonance with quantum numbers $(I,J^P)=(0,0^+)$.
It is located 6.2 MeV below the $\Sigma_c\Sigma_c$ 
threshold and has a width of 4.7 MeV. The foregoing 
two-body state contributes to 
generate a $N \Sigma_c\Sigma_c$ resonance with 
quantum numbers $(I,J^P)=(1/2,1/2^+)$ and a separation 
energy of 0.2 MeV.
\end{abstract}

\maketitle

\section{Introduction}
\label{secI}

The existence of molecules made of heavy baryons is a hot topic
in nowadays hadronic physics~\cite{Lee11,Meg11,Liz12,Oka13,Hua14,Men17,Yan20}.
The observation in 2017 by the LHCb Collaboration of a doubly charmed
baryon~\cite{Aai17} increased the interest in exotic states 
containing pairs of charmed quarks. Right now, the LHCb Collaboration
has reported two structures matching the lineshape of a resonance just 
above twice the $J/\Psi$ mass, that could
originate from a hadron containing two charm quarks~\cite{Aai20}. 
Although the existence of exotic structures containing pairs of
heavy quarks is a long-term challenge~\cite{Ade82},
it has recently been noticed, for example in Refs.~\cite{Kar17,Eic17,Her20},
that doubly charmed tetraquarks are the first to be at the edge of binding.

On general grounds, the main motivation to wonder about the existence
of heavy-baryon molecules is rooted in the reduction of 
the kinetic energy due to larger reduced masses. 
However, such molecular states could be the concatenation of
several effects and not just a fairly attractive interaction. The coupling 
between nearby channels, conflicts between different terms of 
the interaction, and non-central forces often 
play a significant role. Some of these contributions 
may be reinforced by the presence of heavy quarks while others 
may become weaker~\cite{Jun19,Ric20}. 

Behind all this there is the understanding of the hadron-hadron interaction
governed by the dynamics of quarks and gluons, which is a topical issue. 
To encourage new experiments and analysis of existing data
it is essential to have detailed theoretical investigations. 
Despite some uncertainty in contemporary interaction 
models, the possible existence of bound states or resonances 
is a key element, because their signs might be clear enough 
to be identified in the experimental data~\cite{Aai20}. 
Thus, it is the purpose of this work to study the possible 
existence of hadronic molecules
or resonances in two- and three-baryon systems with two
units of charm, in particular, $\Lambda_c\Lambda_c$, $\Sigma_c\Sigma_c$,
$N\Lambda_c\Lambda_c$ and $N\Sigma_c\Sigma_c$ states.
When tackling this problem, one has to manage with an important 
difficulty, namely the complete lack of experimental data. Therefore, 
the generalization of models describing two-hadron interactions 
in the light-flavor sector could offer 
insight about the unknown interaction of hadrons with heavy flavors.
 
Following these ideas, we will make use of a constituent quark
model (CQM) tuned on the description of the $NN$ interaction~\cite{Val05} as 
well as the meson~\cite{Vij05} and baryon~\cite{Vag05,Val08} 
spectra in all flavor sectors,
to obtain parameter-free predictions that will hopefully be testable in 
future experiments. Let us note that the study of 
the interaction between charmed baryons has become an interesting subject 
in several contexts~\cite{Aai20,Wie11,Hoh11,Nou17,Fuj17} and 
it may shed light on the possible 
existence of exotic nuclei with heavy 
flavors~\cite{Dov77,Gar15,Mae16,Hos17,Kre18,Miy18}. 

The paper is organized as follows. In Sec.~\ref{secII} we describe
and analyze particular aspects of the $S$ wave two-body subsystems: 
$N\Lambda_c$, $N\Sigma_c$, $\Lambda_c\Lambda_c$ and $\Sigma_c\Sigma_c$.
Section~\ref{secIII} is devoted to the study of the lightest
$N\Lambda_c\Lambda_c$ and $N\Sigma_c\Sigma_c$ three-body
systems. Finally, in Sec.~\ref{secIV} we summarize our main conclusions.

\section{Two-baryon systems}
\label{secII}
The two-body interactions that are necessary to study the charm $+2$ two- and 
three-baryon systems have been discussed at length in the 
literature~\cite{Gar19,Car15}. They are derived from the
CQM~\cite{Val05,Vij05,Vag05,Val08}. The capability of the model is endorsed
by the nice description of the $NN$ phase shifts, 
as can be seen in Figs. 2, 3 and 4 of 
Ref.~\cite{Gar99}. The $N\Lambda_c$ and $N\Sigma_c$ 
interactions have been presented 
and discussed in detail in Ref.~\cite{Gar19}, in comparison with 
the other approaches available in the literature, 
in particular recent lattice QCD studies~\cite{Miy18}. 
The $\Lambda_c\Lambda_c$ and $\Sigma_c\Sigma_c$ interactions 
have been consistently derived within the CQM in Ref.~\cite{Car15},
also in comparison with the alternative approaches available
in the literature. 
We refer the reader to Refs.~\cite{Gar19,Car15} for a thorough
description of the derivation and analysis of the two-body interactions.
As can be seen in Table 1 of Ref.~\cite{Gar19} and Table II
of Ref.~\cite{Car15} all two-body interactions are consistently
derived with the same set of parameters.
In the following we highlight some peculiarities of the
two-body interactions that are
relevant to the purpose of the present work.

\begin{figure}[b]
\vspace*{-0.5cm}
\includegraphics[width=.6\columnwidth]{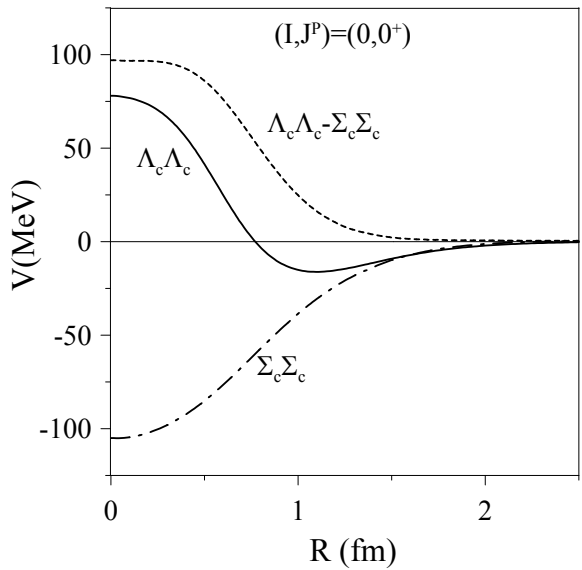}
\vspace*{-6.cm}
\caption{Charm $+2$ $(I,J^P)=(0,0^+)$ two-body interactions.}
\label{fig1}
\end{figure}

We summarize in Table~\ref{tab1} the low-energy parameters of the two-body
systems in the charm $+1$ and $+2$ sectors. The scattering length becomes
complex for those two-body channels with open lower mass two-body states.
The two-body interactions are in general attractive
but not sufficient for having bound states,
in agreement with lattice QCD estimations~\cite{Miy18}.
The singlet isospin $1/2$ and triplet isospin $3/2$ $\Sigma_c N$ 
interactions are the only repulsive ones.
The last line of Table~\ref{tab1} presents the results for the uncoupled 
$\Sigma_c\Sigma_c$ isosinglet system~\footnote{Note that all other
$\Sigma_c\Sigma_c$ $S$ wave states are clearly unbound,
see Fig. 6 of Ref.~\cite{Car15}.}.
It can be seen how the scattering length is positive and larger 
than the range of the interaction, see Fig.~\ref{fig1}, pointing to existence
of a bound state that will be discussed further below. 
\begin{table}[t]
\caption{CQM results for the $^1S_0$ and $^3S_1$ scattering lengths ($\mathrm{a}_s$ and $\mathrm{a}_t$) and effective
range parameters ($r_s$ and $r_t$) in fm for the different $S$ wave $Y_c N$ and $Y_cY_c$ systems 
($Y_c = \Lambda_c$ or $\Sigma_c$). The results shown in the last line, marked by a $\dagger$,
correspond to the uncoupled $\Sigma_c\Sigma_c$ system.}
\begin{tabular}{cp{0.5cm}cp{1cm}cp{0.35cm}cp{1cm}cp{0.35cm}c} \hline\hline
$I$                    && System                   && $\mathrm{a}_s$           &&  $r_s$        &&  $\mathrm{a}_t$          &&  $r_t$  \\
\hline
\multirow{2}{*}{$1/2$} &&  $\Lambda_c N$           && $-$0.86                  &&  5.64         &&  $-$2.31                 &&  2.97  \\
                       &&  $\Sigma_c N$            && $0{.}74 - i\, 0{.}18$    &&  $-$          &&  $-5{.}21 - i\, 1{.}96$  &&  $-$   \\   
$3/2$                  &&  $\Sigma_c N$            && $-$1.25                  &&  8.70         &&  0.95                    &&  4.89  \\
\multirow{3}{*}{$0$}   &&  $\Lambda_c \Lambda_c$   && $-$6.45                  &&  2.29         &&  $-$                     &&  $-$   \\ 
                       &&  $\Sigma_c \Sigma_c$     && $-0{.}014 + i\, 0{.}26$  &&  $-$          &&  $-$                     &&  $-$   \\  
                       &&  $\Sigma_c \Sigma_c$$^\dagger$     && 1.79                     &&  0.44         &&  $-$                     &&  $-$   \\ 
\hline
\end{tabular}
\label{tab1}
\end{table}

Of particular interest are the results for the lightest charm $+2$ channel, with quantum numbers
$(I,J^P)=(0,0^+)$. We show in Fig.~\ref{fig1} the two-body potentials involved in this
channel. The $\Lambda_c\Lambda_c$
interaction is slightly attractive at intermediate distances but, however, 
repulsive at short range. It is decoupled from the
closest two-baryon threshold, the $N\Xi_{cc}$ state~\cite{Car15}, 
which is relevant for the possible existence of 
a resonance in the strange sector~\cite{Sas20,Gar20}. 
There is a general agreement on the overall
attractive character of the $\Sigma_c\Sigma_c$ 
interaction~\cite{Hua14,Lee11,Meg11}. 
Finally, the CQM coupling between the $\Lambda_c\Lambda_c$ 
and $\Sigma_c\Sigma_c$ channels is a
bit stronger than in hadronic theories, based solely 
on a one-pion exchange potential~\cite{Meg11}, due
to quark-exchange effects~\cite{Car15}.
All of this fits
the scenario of the strange sector, as can be seen by comparing with
Fig. 1(b) of Ref.~\cite{Car12}, but the absence of the one-kaon
exchange potential gives rise to a less attractive interaction. 
\begin{figure}[t]
\vspace*{-1.0cm}
\includegraphics[width=.6\columnwidth]{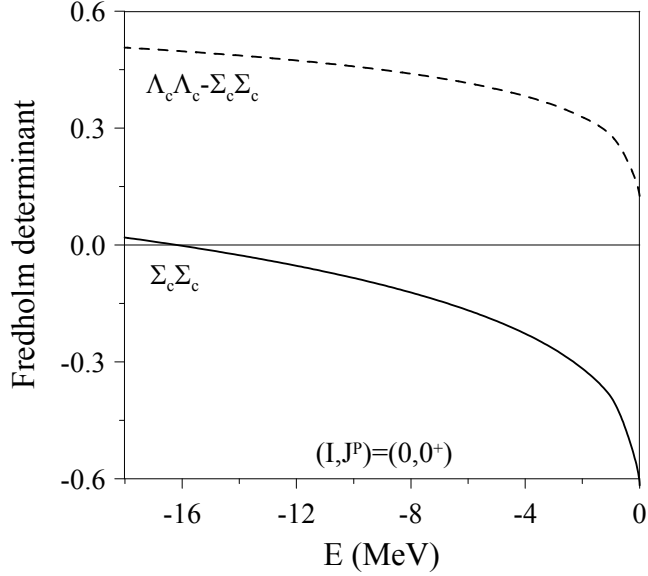}
\vspace*{-6.cm}
\caption{Fredholm determinant of the two-body $(I,J^P)=(0,0^+)$ charm $+2$ channel. The dashed line
corresponds to the $\Lambda_c\Lambda_c - \Sigma_c\Sigma_c$ coupled system, whereas the
solid line considers only the $\Sigma_c\Sigma_c$ channel. The zero energy 
represents the mass of the lowest threshold, $2m_{\Lambda_c}$ for the 
dashed line and $2m_{\Sigma_c}$ for the solid line.}
\label{fig2}
\end{figure}

In Fig.~\ref{fig2} we present the Fredholm determinant~\cite{Gar87} for the 
two-body $(I,J^P)=(0,0^+)$ charm $+2$ channel
in two different cases. The dashed line corresponds to the 
result considering the full
coupling between the $\Lambda_c\Lambda_c$ and  $\Sigma_c\Sigma_c$ states, 
whereas the solid line considers only the $\Sigma_c\Sigma_c$ channel.
The coupled channel calculation shows an attractive character
but not sufficient to generate a bound state, the Fredholm determinant
does not become negative for energies below threshold. This result is in agreement
with other estimations in the literature~\cite{Hua14,Lee11,Meg11} in which
in spite of the attractive character of the $\Lambda_c\Lambda_c$ interaction, 
the central potential alone is not enough to generate a bound state. 
The coupling to larger mass channels could be important for the
existence of a bound state or a resonance. However, due to
the large mass difference between the two coupled channels in 
the $(I,J^P)=(0,0^+)$ partial wave, 338 MeV, the 
coupled channel effect is weakened.
Let us just note that, for example, in the strange sector 
the coupling to the $N\Sigma$ state is relevant for 
the $N\Lambda$ system~\cite{Gar07}
due to a smaller mass difference, 
$M(\Sigma) - M(\Lambda)=$ 77 MeV.
Heavier mass channels play a minor role, 
such as the $\Delta\Delta$ channel (584 MeV above
the $NN$ threshold)
in the $NN$ system~\cite{Val95}. Thus,
one does not expect higher channels, 
as it could be $\Sigma_c^*\Sigma_c^*$ (468 MeV
above the $\Lambda_c\Lambda_c$ threshold) 
to play a relevant role, as it has been explicitly checked in
the literature~\cite{Hua14}. 

Due to the large mass difference between the $\Lambda_c\Lambda_c$ 
and $\Sigma_c\Sigma_c$ channels, we have studied the 
uncoupled $\Sigma_c\Sigma_c$ system. 
The dynamics could be dominated by the attraction 
in the $\Sigma_c\Sigma_c$ channel in a way that the $\Lambda_c\Lambda_c$
channel would be mainly a tool for detection. This mechanism is somewhat 
related to the 'synchronization of resonances' proposed 
by D.~Bugg~\cite{Bug08}. A similar situation 
could be the case of the $d^*(2380)$ resonance in the 
$\Delta\Delta$ system, see Ref.~\cite{Don18} for a recent review.
The result is depicted in Fig.~\ref{fig2} by the solid line,
showing a bound state 16.2 MeV below the $\Sigma_c\Sigma_c$ threshold,
corresponding to the scattering length given in the last 
line of Table~\ref{tab1}. However, since the $\Lambda_c\Lambda_c$ 
channel is open, the $\Sigma_c\Sigma_c$ state would decay 
showing a resonance behavior. 
This scenario, a two-body coupled channel problem showing a 
bound state in the upper channel but not in the coupled channel
calculation has been studied in detail in Ref.~\cite{Gar18}.
It was demonstrated how the width of the resonance does not 
come only determined 
by the available phase space for its decay to the detection channel, 
but it greatly depends on the relative position of the mass of the 
resonance with respect to the masses of the coupled-channels 
generating the state.~\footnote{The equivalence of the results obtained
using a two-cluster interaction or a variational approach for the
multiquark problem has been recently shown, see for example in Ref.~\cite{Car19}.
Dealing with resonances, the two-cluster interaction allows to look for
the poles of the propagator without resorting to numerical
extensions of the variational approach, like the complex scaling method,
that would just give an indication about the possible existence of a resonance.}

Thus, making use of the interactions given in Fig.~\ref{fig1},
we have studied the width of the resonance 
produced in between the two thresholds, $\Lambda_c\Lambda_c$ and $\Sigma_c\Sigma_c$.
The Lippmann-Schwinger equation 
in the case of $S$-wave interactions is  written as,
\begin{eqnarray}
t^{ij}(p,p';E) &=& V^{ij}(p,p')+\sum_{k=1,2} \int_0^\infty {p^{\prime\prime}}^2
dp^{\prime\prime} \nonumber \\ & \times & \!\!
V^{ik}(p,p^{\prime\prime})\,
\frac{1}{E-\Delta E \,\, \delta_{2,k}-{p^{\prime\prime}}^2/2\mu_k+i\epsilon}\, 
t^{kj}(p^{\prime\prime},p';E)\, ,
\,\,\,\,\,\,\,\, i,j =1,2,
\label{eq1} 
\end{eqnarray}
where $\mu_1=m_{\Lambda_c}/2$, $\mu_2=m_{\Sigma_c}/2$, 
and $\Delta E=2 m_{\Sigma_c} - 2 m_{\Lambda_c}$.
The interactions in momentum space are given by,
\begin{equation}
V^{ij}(p,p')=\frac{2}{\pi}\int_0^\infty r^2dr\; j_0(pr)V^{ij}(r)j_0(p'r) \, ,
\label{eq2} 
\end{equation}
where $V^{ij}(r)$ are the two-body potentials in Fig.~\ref{fig1}.
The resonance exists at an energy $E=E_R$ such that the phase 
shift $\delta(E_R)=90^\circ$, for energies between the 
$\Lambda_c\Lambda_c$ and $\Sigma_c\Sigma_c$ thresholds, i.e., 
$0 < E_R < \Delta E$. The mass of the 
resonance is given by $W_R=E_R + 2 m_{\Lambda_c}$.
The width of the resonance
is calculated using the Breit-Wigner formula as~\cite{Bre36,Cec14,Cec08},
\begin{equation}
\Gamma (E) =\lim\limits_{E \to E_R}\, \frac{2(E_R-E)}{\text{cotg}[\delta(E)]} \, .
\label{eq4} 
\end{equation}
Although the Breit-Wigner 
formula is not very accurate close to threshold, however, we have explicitly 
checked by analytic continuation of the S matrix on the second Riemann sheet 
that at low energy the width follows the expected $\Gamma \sim E^{1/2}$ 
behavior.

Using the formalism described above we have calculated the width of the 
$\Sigma_c\Sigma_c$ state. We found a resonance
331.8 MeV above the $\Lambda_c\Lambda_c$ threshold, 6.2 MeV below the
$\Sigma_c\Sigma_c$ threshold, with a width of 4.7 MeV. 
As a consequence of the coupling to the lower $\Lambda_c\Lambda_c$
channel, the pole approaches the threshold moving from 
$-16.2$ MeV in the real axis to the complex plane, $(-6.2 - i \, 4.7/2)$ MeV.
The mechanism we have discussed helps in understanding the narrow width of 
experimental resonances found in the heavy hadron spectra with a large phase
space in the decay channel. The observation of a small width for the decay 
to a low-lying channel could thus point to a dominant contribution of 
some upper channel to the formation of the resonance.

\section{The three-baryon system}
\label{secIII}

The $\Lambda_c\Lambda_c - \Sigma_c\Sigma_c$ system in a pure $S$ wave configuration has 
quantum numbers $(I,J^P)=(0,0^+)$ so that adding one more nucleon, the
$N\Lambda_c\Lambda_c$ system has necessarily quantum numbers $(I,J^P)=(1/2,1/2^+)$.
It is coupled to the $N\Sigma_c\Sigma_c$ channel.
A detailed description of the Faddeev equations of the 
three-body system can be found in Ref.~\cite{Gar14}. 
It has been explained how to deal with coupled channels
containing identical particles of various types in the upper
and lower channels. We show in Table~\ref{tab2}
the different two-body channels that contribute to the
$N\Lambda_c\Lambda_c - N\Sigma_c\Sigma_c$ 
$(I,J^P)=(\frac{1}{2},\frac{1}{2}^+)$ three-body system.
Notice that the charm $+2$ $S$ wave channels 
$\Lambda_c\Sigma_c$ and $\Sigma_c\Sigma_c$ with isospin 1
are not considered since they do not couple to the 
isosinglet $\Lambda_c\Lambda_c$
two-body subsystem. Therefore, the Faddeev equations
of the $(I,J^P)=(1/2,1/2^+)$ three-body system are of the form,
\begin{eqnarray}
T_{N\Lambda_c}^{\Lambda_c} = &&
-t_{N\Lambda_c}^{\Lambda_c} G_0 T_{N\Lambda_c}^{\Lambda_c}
+2 \, t_{N\Lambda_c}^{\Lambda_c} G_0T_{\Lambda_c\Lambda_c}^N
+ t_{N\Lambda_c-N\Sigma_c}^{\Lambda_c} G_0T_{N\Lambda_c}^{\Sigma_c}
\nonumber \\ 
T_{N\Lambda_c}^{\Sigma_c} = &&
t_{N\Lambda_c}^{\Sigma_c} G_0T_{N\Sigma_c}^{\Lambda_c}
+2 \, t_{N\Lambda_c-N\Sigma_c}^{\Sigma_c} G_0T_{\Sigma_c\Sigma_c}^N
+ t_{N\Lambda_c-N\Sigma_c}^{\Sigma_c} G_0T_{N\Sigma_c}^{\Sigma_c}
\nonumber \\ 
T_{N\Sigma_c}^{\Lambda_c} = &&
t_{N\Sigma_c}^{\Lambda_c} G_0T_{N\Lambda_c}^{\Sigma_c}
+2 \, t_{N\Sigma_c-N\Lambda_c}^{\Lambda_c} G_0T_{\Lambda_c\Lambda_c}^N
+ t_{N\Sigma_c-N\Lambda_c}^{\Lambda_c} G_0T_{N\Lambda_c}^{\Lambda_c}
\nonumber \\ 
T_{N\Sigma_c}^{\Sigma_c} = &&
-t_{N\Sigma_c}^{\Sigma_c} G_0T_{N\Sigma_c}^{\Sigma_c}
+2 \, t_{N\Sigma_c}^{\Sigma_c} G_0T_{\Sigma_c\Sigma_c}^N
+ t_{N\Sigma_c-N\Lambda_c}^{\Sigma_c} G_0T_{N\Sigma_c}^{\Lambda_c}
\nonumber
\end{eqnarray}
\begin{eqnarray}
T_{\Lambda_c\Lambda_c}^N = &&
t_{\Lambda_c\Lambda_c}^N G_0T_{N\Lambda_c}^{\Lambda_c}
+t_{\Lambda_c\Lambda_c-\Sigma_c\Sigma_c}^N G_0T_{N\Sigma_c}^{\Sigma_c}
\nonumber \\
T_{\Sigma_c\Sigma_c}^N = &&
t_{\Sigma_c\Sigma_c}^N G_0T_{N\Sigma_c}^{\Sigma_c}
+t_{\Sigma_c\Sigma_c-\Lambda_c\Lambda_c}^N G_0T_{N\Lambda_c}^{\Lambda_c} \, ,
\label{eq71} 
\end{eqnarray}
where $t_{ij}^k$ are the two-body $t$-matrices that already contain 
the coupling among all two-body channels contributing to a given 
three-body state, see Table~\ref{tab2}. $G_0$ is the propagator of 
three free particles. The superscript of the
amplitudes indicates the spectator
particle and the subscript the interacting pair.
\begin{table}[t]
\caption{$S$ wave two-body channels $(i,j)$ of the various subsystems
that contribute to the  $N\Lambda_c\Lambda_c - N\Sigma_c\Sigma_c$
$(I,J^P)=(\frac{1}{2},\frac{1}{2}^+)$ three-body system.} 
\begin{ruledtabular} 
\begin{tabular}{ccccc} 
& Two-body subsystem  & Spectator & $(i,j)$  & \\
\hline
& $\Lambda_c\Lambda_c$  & $N$         & (0,0)   & \\
& $N\Lambda_c$          & $\Lambda_c$ & $(\frac{1}{2},0)$,$(\frac{1}{2},1)$ & \\
& $\Sigma_c\Sigma_c$    & $N$         & (0,0),(1,1)   & \\
& $N\Lambda_c$          & $\Sigma_c$  & $(\frac{1}{2},0)$,$(\frac{1}{2},1)$ & \\
& $N\Sigma_c$           & $\Lambda_c$ & $(\frac{1}{2},0)$,$(\frac{1}{2},1)$ & \\
& $N\Sigma_c$           & $\Sigma_c$  & $(\frac{1}{2},0)$,$(\frac{1}{2},1)$,   
                                $(\frac{3}{2},0)$,$(\frac{3}{2},1)$ & \\
\end{tabular}
\end{ruledtabular}
\label{tab2} 
\end{table}

\begin{figure*}[t]
\vspace*{-0.5cm}
\includegraphics[width=.60\columnwidth]{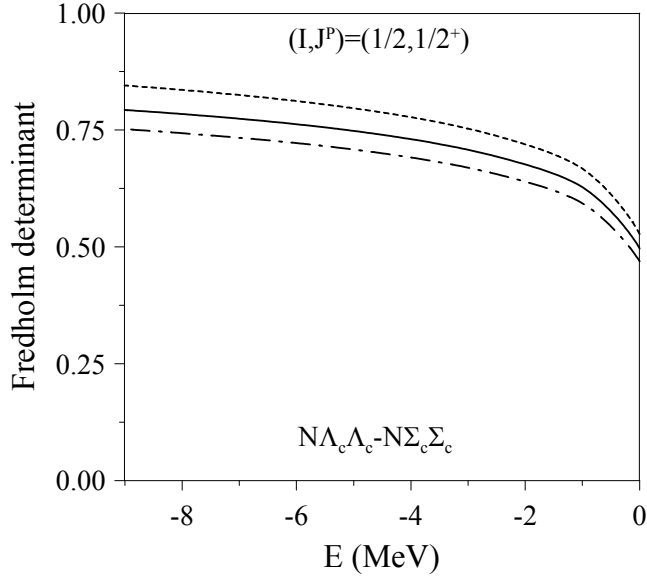}
\vspace*{-6.cm}
\caption{Fredholm determinant of the  $N\Lambda_c\Lambda_c - N\Sigma_c\Sigma_c$
$(I,J^P)=(\frac{1}{2},\frac{1}{2}^+)$ three-body state. 
See text for details.}
\label{fig3}
\end{figure*}

The results are presented in Figure~\ref{fig3}.
We have performed three different calculations. First,
we have included the three-body amplitudes 
of the first two lines of Table~\ref{tab2} that do not 
contain the $\Sigma_c$ baryon, the result being 
depicted by the dotted line.
As it could have been expected, there exists attraction
due to the attractive character of the $N\Lambda_c$ 
and $\Lambda_c\Lambda_c$ interactions discussed in Sect.~\ref{secII}.
However, the attraction is not sufficient for having a
bound state. Then, we have included the amplitudes containing the
$\Sigma_c\Sigma_c$ two-body subsystem, third line in Table~\ref{tab2},
and all isospin 1/2 three-body amplitudes containing a $\Sigma_c$ baryon
either as spectator or as a member of the interacting pair,
lines 4 to 6 of Table~\ref{tab2}.
The result corresponds to the dashed-dotted line.
The coupled channel effect induces some additional attraction, 
reducing the value of the Fredholm determinant. 
Finally, we added the isospin $3/2$ $N\Sigma_c$ amplitudes indicated
in the last line of Table~\ref{tab2}, the result is 
depicted by the solid line. 
Although the singlet amplitudes increase the attraction, the repulsive 
triplet isospin $3/2$ $N\Sigma_c$ amplitude, discussed in Sect.~\ref{secII},
induces an overall repulsion in the three-body system, increasing 
the value of the Fredholm determinant.
\begin{figure*}[t]
\vspace*{-0.5cm}
\includegraphics[width=.60\columnwidth]{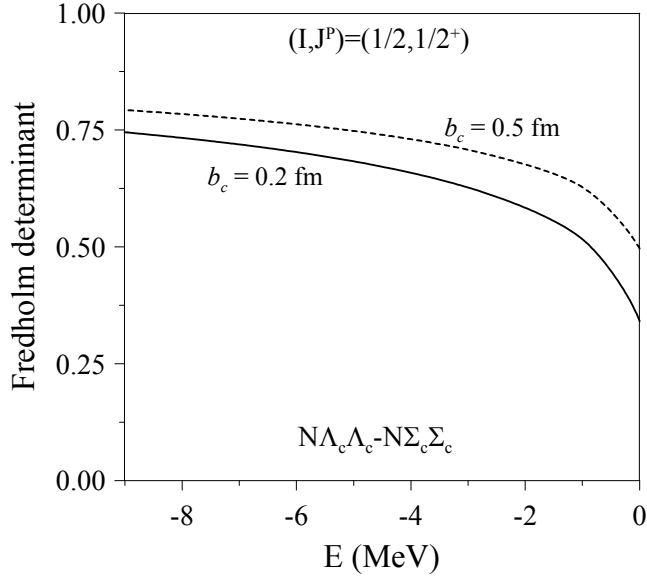}
\vspace*{-6.cm}
\caption{Fredholm determinant of the  $N\Lambda_c\Lambda_c - N\Sigma_c\Sigma_c$
$(I,J^P)=(\frac{1}{2},\frac{1}{2}^+)$ three-body state for two different
values of the Gaussian size parameter of the charm quark 
wave function, $b_c$.} 
\label{fig4}
\end{figure*}

We have studied the dependence of the results
on the only free parameter of the interacting potential, the Gaussian size
parameter of the charm quark wave function, $b_c$. It has been illustrated in
Fig. 4 of Ref.~\cite{Gar19} how the central $Y_c N$ interactions become 
more attractive for smaller values of $b_c$. However, the weakening of the 
non-central contributions generates a slightly less attractive systems
in the presence of coupled-channel effects, see right panel of Fig. 5
of Ref.~\cite{Car15}. It was argued in Ref.~\cite{Car11}
that the smaller values of $b_c$ are preferred to get consistency with 
calculations based on infinite expansions, as hyperspherical harmonic 
expansions~\cite{Vij09}, where the quark wave function is not postulated. 
This also agrees with simple harmonic oscillator relations 
$b_c=b_n\sqrt{\frac{m_n}{m_c}}$~\footnote{$b_n$ and $m_n$
refer to the light quarks, they are given in Table 1 of Ref.~\cite{Gar19} 
and Table II of Ref.~\cite{Car15}}, leading to the best suited value
$b_c=0.2$ fm for the study of the charm sector~\cite{Car11}. Thus,  
in Fig.~\ref{fig4} we show the Fredholm determinant for two different values
of $b_c$, 0.5 and 0.2 fm. As can be seen, the attraction increases 
for smaller values of $b_c$, the effect not being enough 
to generate a bound state. Such tiny contributions might well be
of importance for states at the edge of binding, 
as discussed in the following.

Guided by the resonance found in the $\Sigma_c\Sigma_c$ system, 
we have finally studied the $N\Sigma_c\Sigma_c$ system without 
coupling to $N\Lambda_c\Lambda_c$,
looking for a three-body resonance. The results are promising if the 
triplet isospin $3/2$ amplitude is not considered. 
Thus, considering only the isospin $1/2$ amplitudes 
we obtain a bound state with a separation energy of 
0.6 MeV. If the singlet isospin
$3/2$ amplitude is included, the separation energy
increases to 0.7 MeV. If the repulsive triplet 
isospin 3/2 $(N\Sigma_c)\Sigma_c$ amplitude is considered, 
the signal of the resonance is lost. However,  
adopting the best suited value for the charm sector, $b_c=0.2$ fm, 
the $N\Sigma_c\Sigma_c$ three-body resonance is still there with a
separation energy of 0.2 MeV.

\section{Summary}
\label{secIV}

In short, we have studied two- and three-baryon systems with two units of charm
looking for possible bound states or resonances. 
All two-baryon interactions are consistently derived 
from a constituent quark model tuned in 
the light-flavor hadron phenomenology.
Our results show a narrow 
two-body resonance with quantum numbers $(I,J^P)=(0,0^+)$.
It is located 6.2 MeV below the $\Sigma_c\Sigma_c$ 
threshold and has a width of 4.7 MeV. 
A detailed study of the coupled $N\Lambda_c\Lambda_c - N\Sigma_c\Sigma_c$
three-body system as well as the uncoupled $N\Sigma_c\Sigma_c$
system shows that the foregoing 
two-body state contributes to 
generate a $N \Sigma_c\Sigma_c$ resonance with 
quantum numbers $(I,J^P)=(1/2,1/2^+)$ and a separation 
energy of 0.2 MeV.

Weakly bound states and resonances are usually very sensitive to potential details and therefore
theoretical investigations with different phenomenological models are highly desirable.
The existence of these states could be scrutinized in the future at the LHC, J-PARC and
RHIC providing a great opportunity for extending our knowledge to some unreached 
part in our matter world.

\section{Acknowledgments}
This work has been partially funded by COFAA-IPN (M\'exico) and 
by Ministerio de Ciencia e Innovaci\'on
and EU FEDER under Contracts No. FPA2016-77177-C2-2-P, 
PID2019-105439GB-C22 and RED2018-102572-T.

\end{document}